\documentclass[onecolumn,pre,a4paper]{revtex4}

\usepackage{bbm}
\usepackage{amsmath}
\usepackage{graphicx}
\usepackage{psfrag}

\newcommand{\bq}{\begin{equation}}
\newcommand{\eq}{\end{equation}}
\newcommand{\ba}{\begin{eqnarray}}
\newcommand{\ea}{\end{eqnarray}}
\newcommand{\eql}[2]{\begin{equation}\label{#1} #2 \end{equation}}
\newcommand{\dd}{\mathrm{d}}
\newcommand{\Det}{\mathrm{Det}}
\newcommand{\D}{\mathrm{D}}
\newcommand{\Tr}{\mathrm{Tr}\ }

\newcommand{\m}{\mathbf}
\newcommand{\X}{\mathbf{X}}
\newcommand{\x}{\mathbf{x}}
\newcommand{\A}{\mathbf{A}}
\newcommand{\C}{\mathbf{C}}
\renewcommand{\O}{\mathbf{O}}
\newcommand{\Q}{\mathbf{Q}}
\newcommand{\1}{\mathbbm{1}}

\newcommand{\bO}{\mathbf{\Omega}}
\newcommand{\bo}{\boldsymbol{\omega}}
\newcommand{\bc}{\mathbf{c}}
\newcommand{\ec}{\mathbf{c}}
\newcommand{\DX}{D\mathbf{X}}

\begin{document}
\title{Spectral properties of empirical covariance matrices for data with power-law tails}
\author{Zdzis\l{}aw Burda}\thanks{burda@th.if.uj.edu.pl}
\author{Andrzej T. G\"orlich}\thanks{atg@th.if.uj.edu.pl}
\author{Bart\l{}omiej Wac\l{}aw}\thanks{Corresponding author:
bwaclaw@th.if.uj.edu.pl}

\affiliation{Mark Kac Center for Complex Systems Research and
Marian Smoluchowski Institute of Physics, \\
Jagellonian University, Reymonta 4, 30-059 Krakow, Poland}

\date{\today}

\begin{abstract}
We present an analytic method for calculating spectral densities
of empirical covariance matrices for correlated data. In this
approach the data is represented as a rectangular random matrix
whose columns correspond to sampled states of the system.
The method is applicable to a class of random matrices with radial
measures including those with heavy (power-law) tails in the
probability distribution. As an example we apply it
to a multivariate Student distribution.
\end{abstract}
\maketitle

\section{Introduction}
Random Matrix Theory provides a useful tool for description
of systems with many degrees of freedom. A large
spectrum of problems in physics \cite{phys}, telecommunication,
information theory \cite{mea,sm1,m,s} and quantitative finance
\cite{lcbp,pea,H6,H7,H8,H9,H10,bj} can be naturally formulated in
terms of random matrices.

In this paper we apply random matrix theory to calculate
the eigenvalue density of the empirical covariance matrix.
Statistical properties of this matrix play an important
role in many empirical applications. More precisely,
the problem which we shall discuss here
can be generally formulated in the following way.
Consider a statistical system with $N$ correlated
random variables. Imagine that we do not know {\em a priori}
correlations between the variables and that we try to learn
about them by sampling the system $T$ times. Results of
the sampling can be stored in a rectangular matrix $\X$
containing empirical data $X_{it}$, where the
indices $i=1,\dots, N$ and $t=1,\dots T$ run over
the set of random variables and measurements, respectively.
If the measurements are uncorrelated in time
the two-point correlation function reads:
\eql{prop1}{
\langle X_{i_1 t_1} X_{i_2 t_2} \rangle = C_{i_1 i_2} \delta_{t_1 t_2}.}
where $\C$ is called correlation matrix
or covariance matrix. For simplicity assume
that $\langle X_{it} \rangle = 0$.
If one does not know $\C$ one can try to reconstruct it
from the data $\X$ using the empirical covariance matrix:
\bq
c_{ij}= \frac{1}{T} \sum_{t=1}^T X_{i t} X_{j t} ,
\label{empir1}
\eq
which is a standard estimator of the correlation matrix.
One can think of $\X$ as of an $N\times T$ random matrix
chosen from the matrix ensemble
with some prescribed probability measure $P(\X) \D\X$.
The empirical covariance matrix:
\bq
\ec=\frac{1}{T} \X \X^\tau
\label{empir2}
\eq
depends thus on $\X$. Here $\X^\tau$ stands for the transpose of $\X$.
For the given random matrix $\X$ the eigenvalue density
of the empirical matrix $\ec$ is:
\eql{rhoxl}{
\rho(\X, \lambda) \equiv \frac{1}{N} \sum_{i=1}^{N}
\delta(\lambda - \lambda_i(\bc)),
}
where $\lambda_i(\bc)$'s denote eigenvalues of
$\m c$. Averaging over all random matrices $\X$:
\eql{rhol}{
\rho(\lambda) \equiv \langle \rho(\X, \lambda) \rangle =
\int \rho(\X, \lambda)\ P(\X)\ \D\X ,}
we can find the eigenvalue density of $\m c$ which is representative
for the whole ensemble of $\X$.
We are interested in how the eigenvalue spectrum of $\m c$
is related to that of $\C$ \cite{M1,M2,M3}.
Clearly, as follows from (\ref{prop1}), the quality
of the information encoded in the empirical covariance matrix $\m c$
depends on the number of samples or more precisely
on the ratio $r=N/T$. Only in the limit $T \rightarrow \infty$, that is
for $r \rightarrow 0$, the
empirical matrix $\m c$ perfectly reproduces the real covariance
matrix $\C$. Recently a lot of effort has been made to understand the
statistical relation between $\m c$ and $\C$ for finite $r$.
This relation plays an important role in the theory
of portfolio selection where $X_{it}$ are identified with normalized stocks'
returns and $\C$ is the covariance matrix for
inter-stock correlations. It is a common practice to
reconstruct the covariance matrix from historical data using
the estimator (\ref{empir1}). Since the estimator is calculated
for a finite historical sample it contains a statistical noise.
The question is how to optimally clean the spectrum of
the empirical matrix $\m c$ from the noise
in order to obtain a best quality estimate of the
spectrum of the underlying exact covariance matrix $\C$.
One can consider a more general problem, where in addition to
the correlations between the degrees of freedom (stocks)
there are also temporal correlations between measurements \cite{bjw}:
\eql{prop2}{
\langle X_{i_1 t_1} X_{i_2 t_2} \rangle = C_{i_1 i_2} A_{t_1 t_2} \ ,}
given by an autocorrelation matrix $\A$.
If $\X$ is a Gaussian random matrix, or more precisely
if the probability measure
$P(\X) \D\X$ is Gaussian, then
the problem is analytically solvable in the limit of large matrices
\cite{bjw,fz,sm3,bgjj}. One can derive then an exact relation
between the eigenvalue spectrum of the empirical covariance matrix
$\ec$ and the spectra of the correlation matrices $\A$ and $\C$.

In this paper we present an analytic solution for a class of
probability measures $P(\X) \D \X$ for which
the marginal distributions of individual degrees of freedom
have power law tails: $p(X_{it}) \sim X_{it}^{-1-\nu}$
which means that the cumulative distribution function falls like $X_{it}^{-\nu}$.
Such kind of distributions has been discussed
previously \cite{rr,llll} but, up to our knowledge,
the spectral density of $\ec$ remained unattainable analytically.
The motivation to study such systems comes from the
empirical observation that stocks' returns on financial
markets undergo non-Gaussian fluctuations with
power-law tails. The observed value of the power-law
exponent $\nu \approx 3$ seems to be universal for a wide class
of financial assets \cite{alpha3a,alpha3b,polish}. Random matrix ensembles
with heavy tails have been recently considered for $0<\nu<2$
using the concept of L\'{e}vy stable distributions
\cite{levy1,levy2,levy3}. Here we will present a method which extrapolates 
also to the case $\nu>2$, being of particular interest for financial
markets.

We will study here a model which on the one hand preserves the
structure of correlations (\ref{prop2}) and on the other hand
has power-law tails in the marginal
probability distributions for individual matrix elements.
More generally, we will calculate the eigenvalue density
of the empirical covariance matrix $\m c$ (\ref{empir2}) for random
matrices $\X$ which have a probability distribution of the form:
\begin{equation}\label{PX}
P_f(\X) \DX = \mathcal{N}^{-1} f(\Tr\ \X^\tau \C^{-1} \X \A^{-1}) \DX,
\end{equation}
where $\DX = \prod_{i,t = 1}^{N,T} \dd X_{i t}$ is a volume element.
The normalization constant $\mathcal{N}$:
\bq
\mathcal{N} = \pi^{d/2} (\Det \C)^{T/2}(\Det \A)^{N/2}  \label{nn}
\eq
and the parameter $d=NT$ have been introduced for convenience.
The function $f$ is an arbitrary non-negative function such that $P(\X)$
is normalized: $\int P(\X) \DX = 1$.

In particular we will consider an ensemble of random
matrices with the probability measure given by a
multivariate Student distribution:
\eql{pnu}{ P_\nu (\X) \D\X =
\frac{\Gamma(\frac{\nu + d}{2})}{\mathcal{N}
\Gamma(\frac{\nu}{2}) \sigma^d }
\left(1 + \frac{1}{\sigma^2}\ \Tr\ \X^\tau \C^{-1}
\X \A^{-1} \right)^{-\frac{\nu + d}{2}} \D\X. }
The two-point correlation
function can be easily calculated for this measure:
\eql{2pfnu}{
\langle X_{i_1 t_1} X_{i_2 t_2} \rangle = \frac{\sigma^2}{\nu -2}
C_{i_1 i_2} A_{t_1 t_2}.}
We see that for $\sigma^2 = \nu \!-\! 2$ and for $\nu>2$
the last equation takes the form (\ref{prop2}). With this
choice of $\sigma^2$ the two-point function becomes independent on $\nu$,
however the formula for the probability measure (\ref{pnu}) breaks down
at $\nu=2$ and cannot be extrapolated
to the range $0 < \nu \le 2$. An alternative and actually
a more conventional choice is $\sigma^2\equiv\nu$
which extrapolates easily to this range. In this case
one has to remember that for $\nu>2$ the exact covariance matrix is
given by $\frac{\nu}{\nu-2} \C$, where $\C$ is the matrix in Eq.
(\ref{pnu}) with $\sigma^2 = \nu$. We will stick to this choice in the
remaining part of the paper.

The marginal probability distribution for a matrix
element $X_{it}$ can be obtained by integrating out
all others degrees of freedom from the probability measure
$P(\X) \D\X$. One can see that for
the Student probability measure (\ref{pnu})
the marginal distributions of individual elements
have by construction power-law tails.
For example if $\C$ is diagonal
$\C=\mathrm{Diag}(C_1^2,\dots,C_N^2)$ and $\A=\1_T$ then
the marginal probability distributions can be found exactly
for each element of the matrix $\X$:
\eql{1ds}{ p_i (X_{i t}) =
\frac{\Gamma(\frac{\nu + 1}{2})}{ \Gamma(\frac{\nu}{2})
\sqrt{\nu \pi} C_i}
\left(1 + \frac{X^2_{i t}}{\nu C_i^2}\right)^{-\frac{\nu + 1}{2}}. }
The distributions $p_i$ fall like $\sim X_{it}^{-1-\nu}$
for large $X_{it}$ with
amplitudes which depend on the index $i$ and are independent of $t$.
If one thinks of a stock market, this means
that stocks' returns have the same tail exponent but different
tail amplitudes. The independence of $t$ means that the distributions
$p_i(X_{it})$ are stationary.
More generally, for any $\C$ and for $\A$ which is
translationally invariant $A_{t_1t_2} = A(|t_1\!-\!t_2|)$
the marginal distributions of entries $X_{it}$
can be shown to have power-law tails with the same exponent $\nu$
for all $X_{it}$ and tail coefficients which depend on $i$
and are independent of $t$, exactly expected from
stocks' returns on a financial market.

The main purpose of this paper is to calculate the spectral
density of the empirical covariance
matrix $\ec$ for the Student distribution (\ref{pnu}).
The method is similar to the one presented in
\cite{lecaer1,lecaer2,tvt,bbp,abul} for a square Hermitian matrix.
It consists in an observation that every quantity averaged
over the probability distribution having the form (\ref{PX})
can be first averaged over $(d\!-\!1)$ ``angular''
variables and then of a ``radial'' variable.
This shall be shortly presented in sections II and III.
In the section IV the main equation for the eigenvalue density
of $\m c$ for the radial ensemble (\ref{PX}) with an arbitrary radial
profile $f$ shall be given.
The section V contains results for the Student
distribution (\ref{pnu}) including some special cases.

\section{Radial measures}

The radial measure (\ref{PX}) depends on one scalar
function $f=f(x^2)$ of a real positive argument.
In this section we shall develop a formalism to calculate
the eigenvalue spectrum $\rho_f(\lambda)$
of the empirical covariance matrix (\ref{empir2})
for such radial ensembles. The calculation
can be simplified by noticing that the dependence
of $\rho_f(\lambda)$ on the matrices $\C$ and $\A$ actually
reduces to a dependence on their spectra.
This follows from an observation that
for a radial measure (\ref{PX}) the integral (\ref{rhol})
defining the eigenvalue density is
invariant under simultaneous transformations:
\ba
\C \rightarrow \tilde{\C} = \O \C \O^\tau \nonumber \\
\A \rightarrow \tilde{\A} = \Q^\tau \A \Q \\
\X \rightarrow \tilde{\X} = \O \X \Q^\tau \nonumber
\ea
where $\O,\Q$ are orthogonal matrices of size
$N \times N$ and $T \times T$, respectively.
Choosing the orthogonal transformations $\O$ and $\Q$
in such a way that $\tilde{\C}$ and
$\tilde{\A}$ become diagonal:
$\tilde{\C}=\mathrm{Diag}(C^2_1,\dots,
C^2_N),\ \tilde{\A}=\mathrm{Diag}(A^2_1,
\dots,A^2_T)$ with all $C_i$'s and $A_t$'s being positive,
we see that $\rho_f(\lambda)$ depends on the matrices
$\C$ and $\A$ indeed only through their eigenvalues.
Therefore, for convenience we shall assume
that $\C$ and $\A$ are diagonal from the very beginning.

The radial form of the measure allows one to
determine the dependence of the eigenvalue
density $\rho_f(\lambda)$ on the radial profile $f(x^2)$.
Intuitively, the reason for that stems from the fact
that one can do the integration for the radial ensembles (\ref{PX})
in two steps: the first step is a sort of angular integration which
is done for fixed $x$ and thus is independent of
the radial profile $f(x^2)$, and the second one
is an integration over $x$.
A short inspection of the formula (\ref{PX})
tells us that fixed $x$ corresponds to fixed trace:
$\Tr \X^\tau \C^{-1} \X \A^{-1}$, and thus that we should first perform the
integration over the fixed trace ensemble. We shall follow
this intuition below.

Let us define a matrix $\x = \C^{-\frac{1}{2}} \X \A^{-\frac{1}{2}}$.
Since we assumed that $\A$ and $\C$ are diagonal,
$\A^{1/2}$ and $\C^{1/2}$ are also diagonal with elements
being square roots of those for $\A$ and $\C$.
The elements of $\x$ are:
\eql{xX}{x_{i t} \equiv \frac{X_{i t}}{C_i A_t}.}
They can be viewed as components $x_j$, $j=1,\dots, d$ of
a $d$-dimensional Euclidean vector, where the index $j$ is
constructed from $i$ and $t$. The length of this vector is:
\eql{xx}{ x^2 \equiv \sum_{j=1}^d x_j^2 = \sum_{i=1}^{N}
\sum_{t=1}^{T} x^2_{i t} = \Tr \x^\tau \x =
\Tr\ \X^\tau \C^{-1} \X \A^{-1},}
and thus the fixed trace matrices $\X$
are mapped onto a $d$-dimensional sphere of the given radius $x$.
It is convenient to parameterize the $d$-dimensional
vector $\x$ using spherical coordinates
$\x = x\bo$, where $\bo^2\equiv \Tr \bo^\tau \bo = 1$.
We can also use these coordinates to represent the matrix $\X$:
\ba
\label{Xxw}
\X &=& \C^{\frac{1}{2}} \x \A^{\frac{1}{2}} =
x \C^{\frac{1}{2}} \bo \A^{\frac{1}{2}} = x \bO(\bo), \nonumber\\
\bO(\bo) &\equiv& \C^{\frac{1}{2}} \bo \A^{\frac{1}{2}},
\ea
where the definition of the matrix $\bO(\bo)$ is equivalent to
$\Omega_{it} \equiv C_i A_t \omega_{it}$. While $\bo$
gives a point on a unit sphere in $d$-dimensional space,
$\bO(\bo)$ gives a radial projection of this point on a
$d$-dimensional ellipsoid of fixed trace:
\eql{ftr}{ \Tr\ \bO^\tau \C^{-1} \bO \A^{-1} = 1. }

\section{Angular integration}
We are now prepared to do the integration over the angular
variables $\D \bo$. In the spherical
coordinates (\ref{Xxw}) the radial measure
(\ref{PX}) assumes a very simple form:
\eql{r}{ P_f(\X) \D\X = \pi^{-d/2} f(x^2) x^{d-1}\dd x \; \D\bo. }
The normalization factor $\mathcal{N}^{-1}$ from Eq. (\ref{PX})
cancels out.
The spherical coordinates $\X = x \bO(\bo)$ allow
us to write the formula for $\rho_f(\lambda)$ in the form:
\eql{rhoff}{
\rho_f(\lambda) = \pi^{-d/2} \int \rho(\X, \lambda)\ P_f(\X)\ \D\X  =
\pi^{-d/2} \int \D\bo\int_0^{\infty} \rho\left(x \bO(\bo), \lambda\right)
f(x^2) x^{d-1} \dd x . }
Although the integration over the angular
and the radial part cannot be entirely separated, we can
partially decouple $x$ from $\bO$ in the first argument
of $\rho(x\bO(\bo),\lambda)$.
It follows from (\ref{rhoxl}) that the rescaling
$\X \rightarrow \alpha \X$ by a constant gives the relation:
\eql{rescaling}{\rho(\alpha \X, \lambda) =
\alpha^{-2} \rho(\X, \alpha^{-2} \lambda).}
This observation can be used to rewrite the equation (\ref{rhoff})
in a more convenient form:
\eql{rhofff}{
\rho_f(\lambda) = \pi^{-d/2}
\int \D \bo \int_0^\infty \rho\left(\bO(\bo), \frac{\lambda}{x^2}\right)\
\ f(x^2) x^{d-3} \dd x = \frac{2}{\Gamma(d/2)} \int_0^{\infty}
\rho_*\left(\frac{\lambda}{x^2}\right) \ f(x^2) x^{d-3} \dd x ,
}
where $\Gamma(z)$ is the Euler gamma function and
\eql{rhostar}{ \rho_*(\lambda) \equiv
\frac{1}{S_{d}} \int \rho\left(\bO(\bo), \lambda\right) \D \bo .}
Here $S_{d}$ denotes the hyper-surface area of $d$-dimensional
sphere of radius one: $S_{d}=2 \pi^{d/2}/\Gamma(\frac{d}{2})$.
As we shall see below the last expression is
an eigenvalue distribution of the
empirical covariance matrix for the fixed trace ensemble defined
as an ensemble of matrices $\X$ such that
$\Tr \X^\tau \C^{-1} \X \A^{-1} = 1$.
From the structure
of the equation (\ref{rhofff}) it is clear that if
$\rho_*(\lambda)$ is known then
$\rho_f(\lambda)$ can be easily calculated for any radial profile
just by doing one-dimensional integral.
So the question which we face now is how to determine
$\rho_*(\lambda)$ for arbitrary $\C$ and $\A$. We will do
this by a trick. Instead of calculating $\rho_*(\lambda)$
directly from Eq. (\ref{rhostar}),
we will express $\rho_*(\lambda)$ by the corresponding
eigenvalue density $\rho_G(\lambda)$ for a Gaussian ensemble,
whose form is known analytically \cite{bjw,levy3}.
Let us follow this strategy in the next section.

\section{Fixed trace ensemble and Gaussian ensemble}
The probability measure for the fixed trace ensemble is defined as
\eql{pfte}{ P_*(\X) \D\X =
\frac{\Gamma(\frac{d}{2})}{\mathcal{N} }\
\delta\left(\Tr (\X^\tau \C^{-1} \X \A^{-1}) - 1\right)\ \D\X .}
In the spherical coordinates $\bo$ the formula reads:
\[ P_*(\X) \D\X = \frac{2}{S_{d}}\ \delta(x^2 - 1)\ x^{d-1} \dd x\ \D \bo. \]
One can easily check that the integration
$\rho_*(\lambda) = \int \rho(\X, \lambda) P_*(\X) \D\X$
indeed gives (\ref{rhostar}). It is also worth noticing that
the normalization condition for $P_*(\X)$ is fulfilled.
Consider now a Gaussian ensemble:
\bq \label{pgoe}
P_{G}(\X) \D\X \equiv
\mathcal{N}^{-1} f_{G}(\Tr \X^\tau \C^{-1} \X \A^{-1}) \D\X,
\eq
where
\eql{fg}{f_{G} (x^2) = \frac{1}{2^{d/2}} e^{- \frac{1}{2}x^2}, }
for which the spectrum $\rho_G(\lambda)$ is known or more precisely
it can be easily computed numerically
in the thermodynamical limit $N,T\to\infty$ \cite{bjw,epjb,app}.
On the other hand as we learned in the previous section,
the density of eigenvalues of the empirical
covariance matrix $\ec$ can be found
applying Eq. (\ref{rhofff})
to the Gaussian radial profile (\ref{fg}):
\eql{gft1}{
\rho_{G}(\Lambda) = \frac{2^{1-d/2}}{\Gamma(\frac{d}{2})}
\int_0^\infty \rho_*\left(\frac{\Lambda}{x^2}\right)\,
x^{d-3}\, e^{- \frac{1}{2}x^2}\, \dd x. }
Changing the integration variable to $y$: $x^2= d y^2$
and rescaling the spectrum $\rho_G$ by $d$:
$\lambda = \frac{\Lambda}{d}$ we eventually obtain:
\eql{gft2}{
d \rho_G( d \lambda) =
\int_0^\infty \rho_*\left(\frac{\lambda}{y^2}\right)\,
\frac{1}{y^2}\left[
\frac{2^{1-d/2} d^{d/2}}{\Gamma(\frac{d}{2})}
y^{d-1}\, e^{- \frac{1}{2}d\,y^2}\right]\, \dd y.}
One can easily check that the formula in the square brackets
tends to the Dirac delta for large matrices because then $d$ goes to infinity:
\[ \lim_{d \to \infty}
\frac{2^{1-d/2}\ d^{d/2}}{\Gamma(\frac{d}{2})}
y^{d-1}\, e^{- \frac{1}{2}d\, y^2} = \delta(y - 1), \]
and thus the integrand in Eq. (\ref{gft2}) gets localized
around the value $y=1$. Therefore for large $d$ we can make the
following substitution:
\eql{limgoe}{ \rho_*(\lambda) = d\rho_G(d\lambda) .}
Inserting it into Eq. (\ref{rhofff})
and changing the integration variable to $y = \frac{d\lambda}{x^2}$
we finally obtain a central equation of this paper:
\eql{rhocgoe}{
 \rho_f(\lambda) = \frac{d^{d/2}}{\Gamma(d/2)} \lambda^{d/2-1} \
\int_0^\infty \rho_G (y) f\left(\frac{d \lambda}{y}\right) y^{-d/2} \dd y .}
The meaning of this formula is the following: for
any random matrix ensemble
with a radial measure (\ref{PX}) the eigenvalue density
function $\rho_f(\lambda)$ is given by a one-dimensional integral
of a combination of the corresponding Gaussian spectrum $\rho_G(\lambda)$
and the radial profile $f(x)$.
The equation holds in the thermodynamic limit:
$d = NT\rightarrow \infty$ and $r=N/T=\rm{const}$.
Since in this limit we are able to calculate the
spectrum $\rho_G(\lambda)$ for arbitrarily chosen $\A,\C$,
the formula (\ref{rhocgoe}) gives us a powerful tool
for computing spectra of various distributions.
In the next section we shall apply it to
the multivariate Student ensemble (\ref{pnu}).

\section{Multivariate Student ensemble}

The radial profile for the Student ensemble (\ref{pnu}) is:
\eql{fn}{ f(x^2) \equiv f_\nu (x^2) =
\frac{\Gamma(\frac{\nu + d}{2})} {\Gamma(\frac{\nu}{2}) \nu^{d/2} }
\left(1 + \frac{x^2}{\nu}\right)^{-\frac{\nu + d}{2}}. }
We have chosen here the standard convention $\sigma^2 = \nu$ since
we would like to calculate the spectrum $\rho_\nu(\lambda)$ also for $\nu\le 2$
(see the discussion at the
end of the first section). Inserting (\ref{fn}) into
the equation (\ref{rhocgoe}):
\[ \rho_\nu(\lambda) =
\Big(\frac{d}{\nu}\Big)^{d/2}
\frac{\Gamma(\frac{\nu + d}{2})}{ \Gamma(\frac{d}{2}) \Gamma(\frac{\nu}{2})}
\lambda^{d/2-1} \int_0^\infty \rho_G(y)\,
\left(1 + \frac{d \lambda}{\nu y}\right)^{-\frac{\nu + d}{2}}
y^{-\frac{d}{2}}\,\dd y ,\]
and taking the limit $d \rightarrow \infty$:
\[ \lim_{d\to \infty} \Big(\frac{d}{\nu}\Big)^{d/2}
\frac{\Gamma(\frac{\nu + d}{2})}{ \Gamma(\frac{d}{2})}
y^{-\frac{d}{2}} \lambda^{\frac{d}{2}-1}\left(1 +
\frac{d \lambda}{\nu y}\right)^{-\frac{\nu + d}{2}} =
\Big(\frac{\nu}{2}\Big)^{\nu/2} e^{-\frac{\nu y}{2 \lambda}}
y^{\frac{\nu}{2}} \lambda^{-\frac{\nu + 2}{2}},\]
we see that the expression for $\rho_\nu(\lambda)$
simplifies to an expression which is independent of $d$:
\eql{rhonu}{\rho_\nu(\lambda) =
\frac{1}{\Gamma(\frac{\nu}{2})}
\Big(\frac{\nu}{2}\Big)^{\nu/2} \lambda^{-\frac{\nu}{2}-1}
\int_0^\infty \rho_{G}(y)\, e^{-\frac{\nu y}{2 \lambda}}
y^{\frac{\nu}{2}}\, \dd y .}
The formula (\ref{rhonu}) works for all $\nu>0$.
From the last equation we can infer the behavior
of $\rho_\nu(\lambda)$ for large $\lambda$. The function
$\rho_G(y)$ has a compact support \cite{bgjj,epjb,bjw}, therefore
for large $\lambda$ the exponential can be approximated well by $1$.
The function $\rho_\nu(\lambda)$ has thus a long tail:
\bq
\rho_\nu(\lambda) \approx \lambda^{-\frac{\nu}{2}-1}
\cdot \frac{1}{\Gamma(\frac{\nu}{2})}
\Big(\frac{\nu}{2}\Big)^{\nu/2} \int_0^\infty
\rho_{G}(y) y^{\frac{\nu}{2}}\, \dd y,
\label{rhonul}
\eq
where the integral does not depend on $\lambda$.
The exponent $-\nu/2-1$ in the above power-law
depends on the index $\nu$ of the original
Student distribution. The change from the power $\nu$
to the power $\nu/2$ comes about because $\m c$ is
a quadratic combination of $\X$.

The power-law tail in the eigenvalue
distribution (\ref{rhonul}) does not disappear
in the limit of large matrices contrary to the power-law
tails in the eigenvalue distribution for
an ensemble of matrices whose elements are independently
distributed random numbers. For such matrices,
for $\nu>2$, the density $\rho(\lambda)$ falls into
the Gaussian universality class and yields the Wishart
spectrum \cite{bz}. One should remember
that the multivariate Student distribution (\ref{pnu})
discussed here does not describe independent
degrees of freedom even for $\A=\1_T$ and $\C=\1_N$, in which
case the degrees of freedom are ``uncorrelated'' but not
independent.

We have learned that the spectrum is unbounded from above.
Let us now examine the lower limit of the spectrum.
Rewriting Eq. (\ref{rhonu})
in the form:
\bq
\rho_\nu(\lambda) =
\frac{2\nu^{\nu/2}}{\Gamma(\frac{\nu}{2})}
\int_0^\infty \rho_G\left(2x\lambda\right)
e^{-\nu x} x^{\nu/2} \ \dd x,   \label{rhonu2}
\eq
we see that as long as $\lambda>0$ the function $\rho_\nu(\lambda)$
is positive since $\rho_G(x)$ is positive on a
finite support. Thus the function $\rho_\nu(\lambda)$ vanishes
only at $\lambda=0$ and it is positive for any $\lambda>0$.
Contrary to the classical Wishart distribution
for the Gaussian measure,
the spectrum (\ref{rhonu}) spreads over the whole real positive semi-axis.
On the other hand, taking the limit $\nu\to\infty$
of Eq. (\ref{rhonu2}) and using the formula:
\bq
\lim_{\nu \to \infty} \frac{2 \nu^{\nu/2}}{\Gamma(\frac{\nu}{2})}
x^{\nu/2}\, e^{- \nu x} = \delta(x - 1/2),
\eq
we obtain $\rho_{\nu\to\infty}(\lambda) = \rho_G(\lambda)$
as expected, because in this limit the radial profile
$f_\nu(x^2)$ given by Eq. (\ref{fn}) for the Student distribution reduces to
the Gaussian one (\ref{fg}).

\section{Examples}

\begin{figure}
\psfrag{xx}{$\lambda$} \psfrag{yy}{$\rho_\nu(\lambda)$}
\includegraphics[width=10cm]{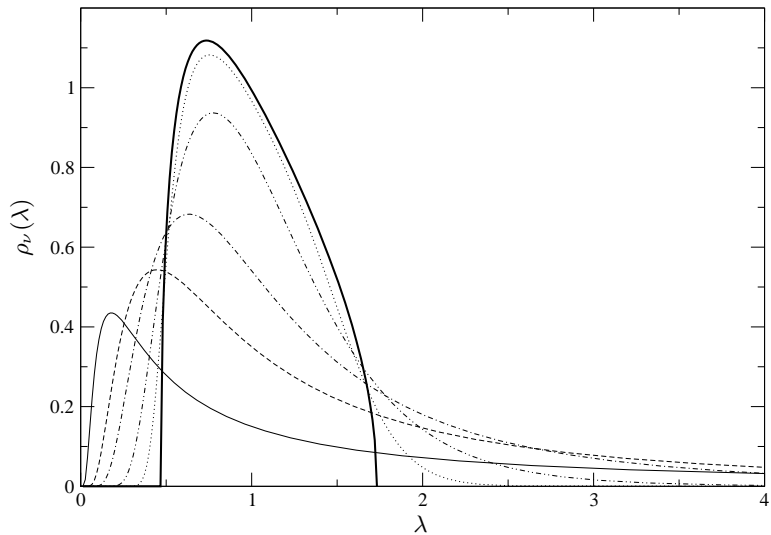}
\caption{Spectra of the covariance matrix $\ec$ for the
Student distribution (\ref{pnu}) with $\C=\1_N$ and $\A=\1_T$, $r=N/T=0.1$,
for $\nu=1/2,2,5,20$ and $100$ (thin lines from solid to dotted),
calculated using the formula (\ref{rhonubk}) and
compared to the uncorrelated Wishart (thick line).
One sees that for $\nu\to\infty$ the spectra tend to
the Wishart distribution.}
\label{fig1}
\end{figure}

\begin{figure}
\vspace{0.5cm}
\psfrag{xx}{$\lambda$} \psfrag{yy}{$\rho_\nu(\lambda)$}
\includegraphics[width=10cm]{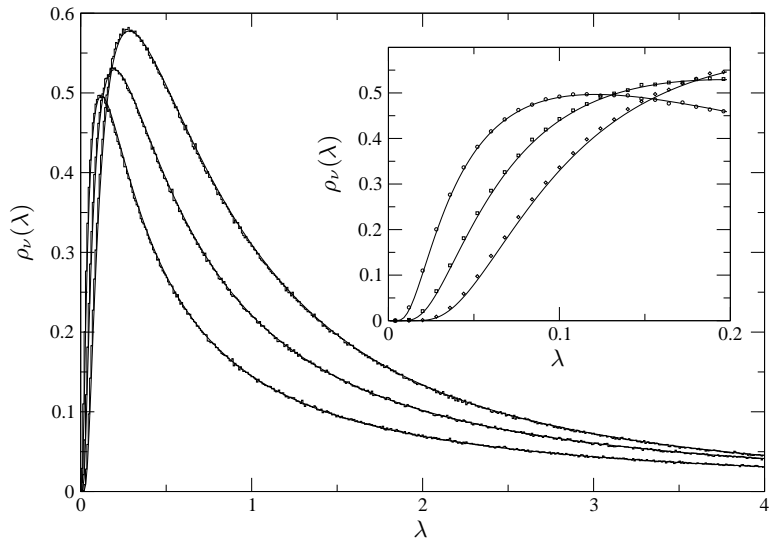}
\caption{Spectra of the empirical
covariance matrix $\ec$ calculated from Eq. (\ref{rhonubk})
with $r=1/3$, compared to experimental data (stair lines)
obtained by the Monte Carlo generation of
finite matrices $N=50, \ T=150$. Inset: the left part of the 
same distributions, points
represent experimental data.}
\label{fig2}
\end{figure}

Let us first consider the case without correlations:
$\C = \1_N$ and $\A = \1_T$. The spectrum of the empirical
covariance for the Gaussian ensemble is given by the Wishart
distribution:
\[ \rho_G(\lambda) =
\frac{1}{2\pi r \lambda}
\sqrt{(\lambda_+ - \lambda) (\lambda - \lambda_-)}, \]
where $\lambda_\pm = (1\pm\sqrt{r})^2$ \cite{M1,M2,M3}.
The corresponding spectrum (\ref{rhonu}) for the Student
ensemble is then: \eql{rhonubk}{
\rho_\nu(\lambda) = \frac{1}{2\pi r \Gamma(\frac{\nu}{2})}
\Big(\frac{\nu}{2}\Big)^{\nu/2} \lambda^{-\nu/2-1}
\int_{\lambda_-}^{\lambda_+} \sqrt{(\lambda_+ - y) (y - \lambda_-)}\,
e^{-\frac{\nu y}{2 \lambda}} y^{\nu/2-1}\, \dd y .}
The integral over $\dd y$ can be easily computed numerically.
Results of this computation for different values of $\nu$
are shown in Fig. \ref{fig1}. For increasing $\nu$
the spectrum $\rho_\nu(\lambda)$ tends to the Wishart distribution
but even for very large $\nu$ it has a tail which touches
$\lambda=0$ as follows from Eq. (\ref{rhonu2}).
In Fig. \ref{fig2} we have plotted $\rho_\nu(\lambda)$ for
$\nu=0.5,1$ and $2$ and compared them
to experimental results obtained
by the Monte-Carlo generation of random matrices drawn from
the corresponding ensemble with the probability measure (\ref{pnu})
for which eigenvalue densities were computed by numerical
diagonalization. The agreement is perfect. Actually it is even
better than for the Gaussian case for the same size $N$. 

\begin{figure}
\psfrag{xx}{$\lambda$} \psfrag{yy}{$\rho_\nu(\lambda)$}
\includegraphics[width=10cm]{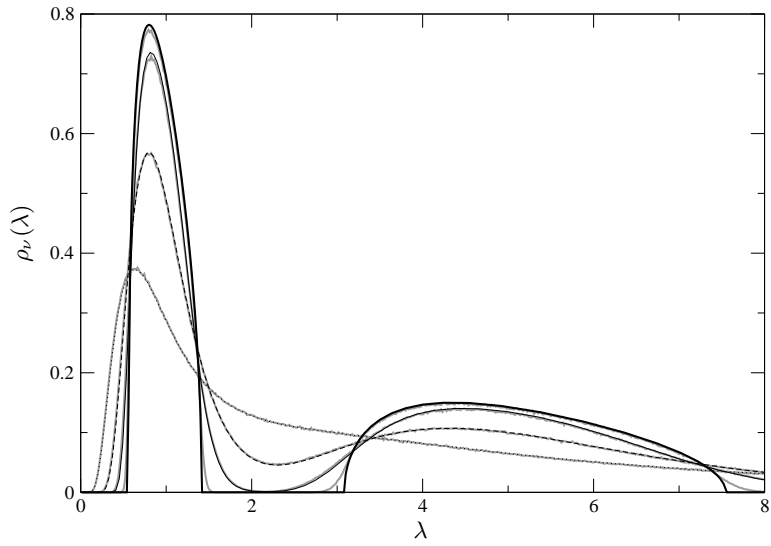}
\caption{Spectra $\rho_\nu(\lambda)$ for $\C$ having two distinct
eigenvalues: $1$ and $\mu$ in proportion $(1-p):p$,
calculated from Eq. (\ref{rhonu}) with $\rho_G$
given by formula (\ref{rho2ww}),
with $r=1/10$, $p=1/2$ and $\mu=5$. Thick solid line corresponds
to the Gaussian case $\nu\to\infty$ while thin lines to
$\nu=5,20,100$. These lines are compared to Monte-Carlo results
obtained by the generation and diagonalization of finite matrices with 
$N=40, \ T=400$ (gray lines), which lie almost exactly on top of them
and can be hardly seen by an unarmed eye.
}
\label{fig3}
\end{figure}

As a second example we consider the case when $\C$ has two
distinct eigenvalues $\lambda_1$ and $\lambda_2$
with degeneracies: $(1-p)N$ for $\lambda_1$ and
$pN$ for $\lambda_2$, where $0 \le p\le 1$.
Such a covariance matrix can be used to model the simplest
effect of sectorization on a stock exchange.
For example if all diagonal elements of the matrix $\C$ are
equal $1$ and all off-diagonal are equal $\rho_0$ ($0<\rho_0<1$)
the model can be used to mimic a collective behavior on the
market \cite{H7,H8}. In this case $\lambda_1=1-\rho_0$ has
a degeneracy $N-1$ and $\lambda_2=1+(N-1)\rho_0$ is non-degenerated,
hence $p=1/N$. The eigenvector corresponding to the larger eigenvalue
$\lambda_2$ can be thought of as describing the correlations of
all stocks. For our purposes it is however more convenient to set
$\lambda_1=1$ and $\lambda_2\equiv\mu$ and $p$ being an arbitrary number
between zero and one.
The corresponding Wishart spectrum $\rho_G(\lambda)$ can be
obtained by solving equations given by a conformal map \cite{bgjj}.
The resulting spectrum has the form:
\bq
\rho_G(\lambda) =
\frac{1}{\pi} \left| \mbox{Im} \frac{M(Z(\lambda))}{\lambda}\right|, \label{rho2ww}
\eq
where
\ba
M(Z) &=& \frac{1-p}{Z-1}+\frac{p\mu}{Z-\mu}, \\
Z(\lambda) &=& -\frac{a}{3}+\frac{(1-i\sqrt{3})(3b-a^2)}{3\cdot 2^{2/3} E}-
\frac{(1+i\sqrt{3})E}{6\cdot 2^{1/3}}, \\
   E &=& \left( 3\sqrt{3} \sqrt{27c^2-18abc+4a^3c+4b^3-a^2b^2}
       -27c+9ab-2a^3\right)^{1/3},
\ea
where $a=r-1-pr-\mu(1-pr)-\lambda$, $b=\lambda(\mu+1)-\mu(1-r)$ and $c=-\lambda\mu$.
Inserting the above formula into Eq. (\ref{rhonu}) we obtain an
integral, which can be computed numerically for arbitrary $r,\mu,p$.
In Fig. \ref{fig3} we show examples of this computation
for different values of the index $\nu$. In the same figure we
compare the analytic results with those obtained by the Monte Carlo
generation and numerical diagonalization of random matrices
for $N=40,T=400$. As before, the agreement between the analytic
and Monte-Carlo results is perfect. We see that the effect on
the spectrum of introducing heavy tails increases with
decreasing $\nu$. When
$\nu$ is decreasing from infinity to zero the two disjoint islands
of the distribution develop a bridge to eventually end up
as a distribution having only one connected component.

\section{Summary}
In the paper we have developed a method for computing spectral densities
of empirical covariance matrices for a wide class of
``quasi-Wishart'' ensembles with radial probability measures.
In particular we have applied this method to determine
the spectral density of the empirical covariance matrix
for heavy tailed data described by a Student multivariate distribution.
We have shown that the spectrum $\rho(\lambda)$ decays 
like $\lambda^{-\nu/2-1}$ where $\nu$ is the index
of Student distribution. The case of $\nu=3$ is of particular 
importance since it can be used in modeling stock markets.
The eigenvalue density spreads over the whole positive semi-axis 
in contrast to the Wishart spectrum which has a finite support.

We have also derived a general formula for the eigenvalue spectrum
of the empirical covariance matrix for radial ensembles. 
The spectrum is given by a one-dimensional integral, which 
can be easily computed numerically. The method works also in the
case of correlated assets.

\section*{Acknowledgements}
We would like to thank Jerzy Jurkiewicz, Maciej A. Nowak,
Gabor Papp and Ismail Zahed for many inspiring discussions.
This work was supported by
Polish Ministry of Science and Information Society Technologies grants:
2P03B-08225 (2003-2006) and 1P03B-04029 (2005-2008)
and EU grants: MTKD-CT-2004-517186 (COCOS)
and MRTN-CT-2004-005616 (ENRAGE).

\end{document}